\documentclass[fleqn,usenatbib,useAMS]{mnras}

\usepackage{graphicx}	
\usepackage{amsmath}	
\usepackage{amssymb}	
\usepackage{multicol}        
\usepackage{bm}		
\usepackage{pdflscape}	
\usepackage{natbib}
\usepackage{times}
\usepackage{calc}
\usepackage{color}
\usepackage{rotating}
\usepackage{lscape}
\bibpunct{(}{)}{;}{a}{}{,} 
\usepackage{longtable}
\usepackage{enumitem} 
\usepackage{hyperref}
\usepackage{pifont}
\usepackage{tikz}
\usepackage[toc,page]{appendix}

\def\oversim#1#2{\lower0.5pt\vbox{\baselineskip0pt \lineskip-0.5pt
     \ialign{$\mathsurround0pt #1\hfil##\hfil$\crcr#2\crcr\sim\crcr}}}

\usepackage[T1]{fontenc}
\usepackage{ae,aecompl}

\usepackage{mathptmx}

\def\aap {{A\&A}}

\def\apjl {{ApJ}}

\def\apjs {{ApJs}}

\title[Properties of PC\,22]{Catching a Grown-Up Starfish Planetary Nebula: I. Morpho-Kinematical study of PC\,22}

\author[L. Sabin et al.] 
{L. Sabin$^{1}$\thanks{E-mail:lsabin@astro.unam.mx (LS)},
M.A.\ G\'omez-Mu\~noz$^{1}$,
M.A.\ Guerrero$^{2}$,
S.\ Zavala$^{3,4}$,
G.\ Ramos-Larios$^{5}$,
\newauthor R.\ V\'azquez$^{1}$,
L.\ Corral$^{5}$,
M.W.\ Blanco C\'ardenas$^{1}$,
P.F.\ Guill\'en$^{1}$,
L.\ Olgu\'{\i}n$^{6}$,
C.\ Morisset$^{7}$,
\newauthor S.\ Navarro$^{5}$  \\  
$^{1}$Instituto de Astronom\'{\i}a, Universidad Nacional Aut\'onoma de M\'exico, Apdo. Postal 877, 22800 Ensenada, B.C., Mexico\\
$^{2}$Instituto de Astrof\'{\i}sica de Andaluc\'{\i}a, IAA-CSIC, C/ Glorieta de la Astronom\'{\i}a s/n, 18008 Granada, Spain\\
$^{3}$Instituto Tecnol\'ogico de Ensenada, Blvd Tecnol\'ogico No. 150, 22780 Ensenada, B. C., Mexico\\
$^{4}$Instituto de Estudios Avanzados de Baja California, A. C., Blvd Tte. Azueta 147, Edif.\ Matematik\'e Planta Baja, 22800 Ensenada, B. C., Mexico \\
$^{5}$Instituto de Astronom\'{\i}a y Meteorolog\'{\i}a, Av.\ Vallarta No.\ 2602, Col.\ Arcos Vallarta, C.P. 44130 Guadalajara, Jalisco, Mexico \\
$^{6}$Dpto.\ de Investigaci\'on en F\'{\i}sica, Universidad de Sonora, Blvd. Rosales-Colosio, Ed. 3H, 83190, Hermosillo, Sonora, Mexico \\
$^{7}$Instituto de Astronom\'{\i}a, Universidad Nacional Aut\'onoma de M\'exico, Apdo. Postal 70264, Mexico D. F. 04510, Mexico}
\date{Last updated 2016 xx xx; in original form 2016 xx xx}
\pubyear{2016}

\begin{document}
\label{firstpage}
\pagerange{\pageref{firstpage}--\pageref{lastpage}}
\maketitle


\begin{abstract}

We present the first part of an investigation on the planetary nebula (PN)
PC\,22 which focuses on the use of deep imaging and high resolution echelle
spectroscopy to perform a detailed morpho-kinematical analysis.
PC\,22 is revealed to be a multipolar PN emitting predominantly in
{[O\,{\sevensize III}]} and displaying multiple non-symmetric outflows.
Its central region is found to be also particularly inhomogeneous with
a series of low ionization structures (knots) located on the path of
the outflows.
The morpho-kinematical model obtained with {\sc SHAPE} indicates that
i) the de-projected velocities of the outflows are rather large, $\geq$100
km\,s$^{-1}$, while the central region has expansion velocities in the
range $\sim$25 to $\sim$45 km\,s$^{-1}$ following the ``Wilson effect'',
ii) the majority of the measured structures share similar inclination,
$\simeq$100\degr, i.e.\ they are coplanar, and
iii) all outflows and lobes are coeval (within the uncertainties). All these results make us to suggest that PC\,22 is an {\it evolved starfish PN}.
We propose that the mechanism responsible for the morphology of PC\,22
consists of a wind-shell interaction, where the fast post-AGB wind flows
through a filamentary AGB shell with some large voids.
  
\end{abstract}


\section{Introduction}

Planetary Nebulae (PNe) are well known for displaying different degrees
of morphological complexity probing their, often intense, internal
dynamical activity ({\it see the imaging catalogues by}, \citealt{Parker2006,Parker2016}; \citealt{Sahai2011} and \citealt{Sabin2014}).
Hence, the presence of outflows, jets, rings or knots each indicate
particular physical processes whose origins are not always well
constrained.
By establishing with accuracy the morpho-kinematics and chemical
characteristics of PNe, one can expect to better understand their
formation history. 
Such multi-approach analyses have been performed to study in great detail
various PNe such as NGC\,3132 \citep{Monteiro2000}, IPHASX\,J194359.5$+$170901
\citep{Corradi2011}, K\,4-55 and Kn\,26 \citep[respectively]{Guerrero1996,
Guerrero2013}, M\,2-48 \citep{lopez-martin2002}, ETHOS\,1
\citep{Miszalski2011}, and NGC\,6309 \citep{Vazquez2008}.  
With this aim in mind, we present in this article a detailed investigation
of a peculiar PN, PC\,22 (PN\,G051.0$-$04.5) located at $\alpha_{2000}$=19:42:03.50, $\delta_{2000}$=+13:50:37.33.

Discovered by \citet{Apriamashvili1959} and later catalogued by
\citet{Peimbert1961}, this object, has not been subject of any extensive targeted study so
far.  
\citet{Manchado1996} classified PC\,22 as an elliptical planetary nebula
with ``internal structures'' (no description of these were made).
Although no binary system has been detected, \citet{Soker1997} predicted
a high probability for the progenitor to have gone through a common envelope with a sub-stellar
companion based on this morphology.
Except for photometric measurements realized at various wavelengths we
found no other relevant piece of information about the PN.

The distance to PC\,22 seems to be relatively well constrained.
\citet{Tajitsu1998} obtained a distance of 5.5 kpc for this object by
fitting a blackbody to the flux of the four IRAS bands, whereas
\citet{Giammanco2011} estimated a distance of 4.0$\pm$0.5 kpc based on
the distance-extinction method built with the first IPHAS data
release\footnote{
Since then a new photometric calibration has been performed by
\citet{Barentsen2014}.
}
. More recently \citet{Frew2016}, using a new H$\alpha$ surface
brightness-radius relation, derived a mean distance of 5.27$\pm$1.52 kpc
adopting log(S(H$\alpha$))= --2.77$\pm$0.10 erg~cm$^{-2}$~s$^{-1}$~sr$^{-1}$.

In this paper we present new deep morpho-kinematical observations which
reveal a wealth of information on PC\,22 and assert its physical structure.
The article is organized as the following:
the observations are described in \S\ref{sec_obs};
the results obtained from each observing technique are described in
\S\ref{sec_res} and this includes our proposed kinematical model.
Finally we conclude our investigation presenting the main findings in \S\ref{sec_find} followed by the discussion and concluding remarks in \S\ref{sec_dis}.
This is the first paper on a series which focuses on the morpho-kinematical
analysis of PC\,22.
We will present a comprehensive chemical analysis in a forthcoming paper
(Sabin et al.\ in preparation).

\section[]{Observations}\label{sec_obs}

\subsection{ALFOSC optical imaging}

\begin{figure*}
\begin{center}
\includegraphics[width=0.8\textwidth]{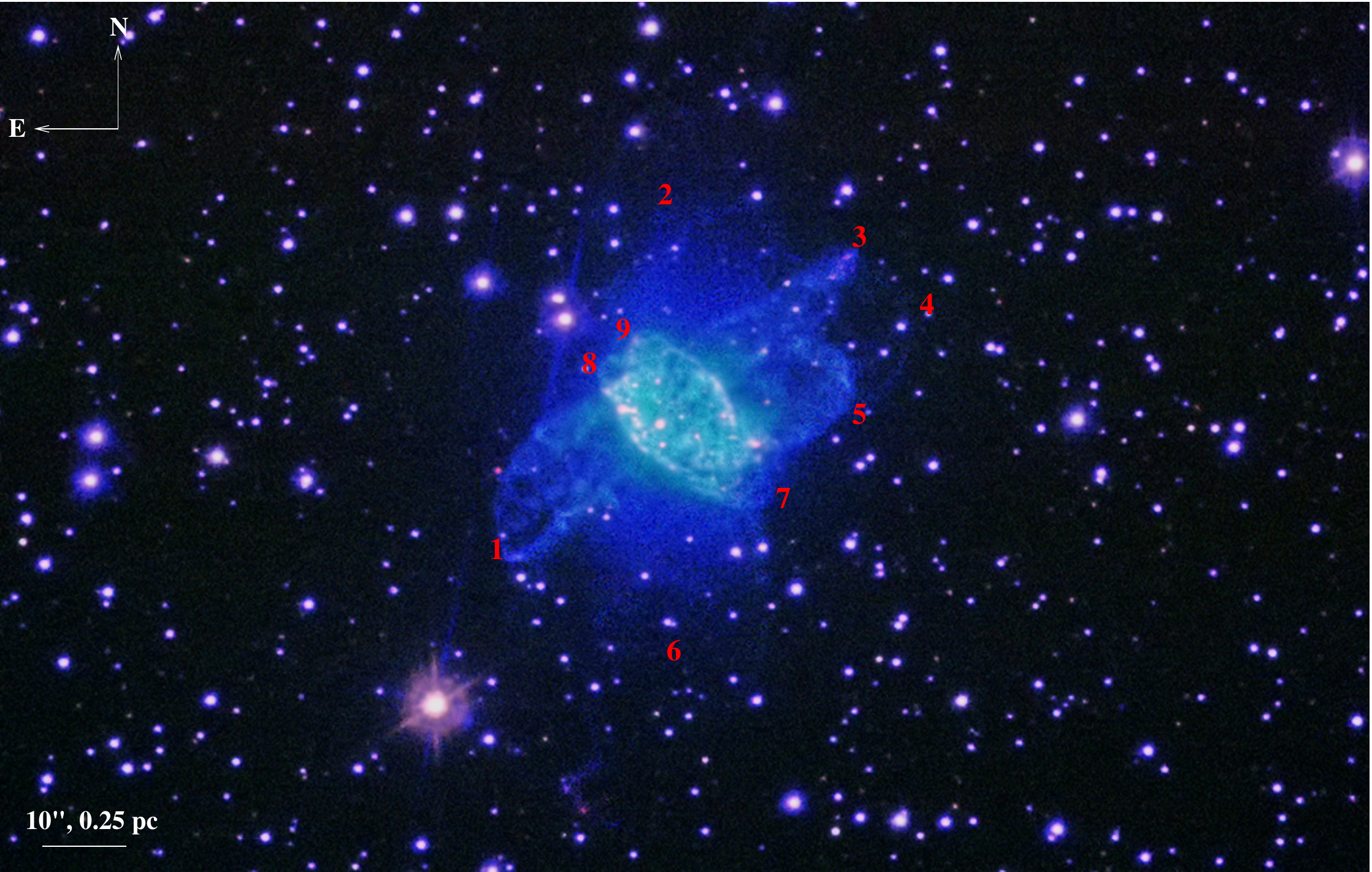}
\caption{
Nordic Optical Telescope RGB picture of PC\,22 (R=H$\alpha$, G=[N~{\sc ii}],
B=[O~{\sc iii}]).
The image reveals a non-homogeneous central elliptical structure and
several ejections reaching distances up to $\sim$50\arcsec\ from the
central star.
The numbers indicate the different outflows identified in this nebula.
}
\label{RGB}
\end{center}
\end{figure*} 

\begin{figure*}
\begin{center}
\includegraphics[width=0.8\textwidth]{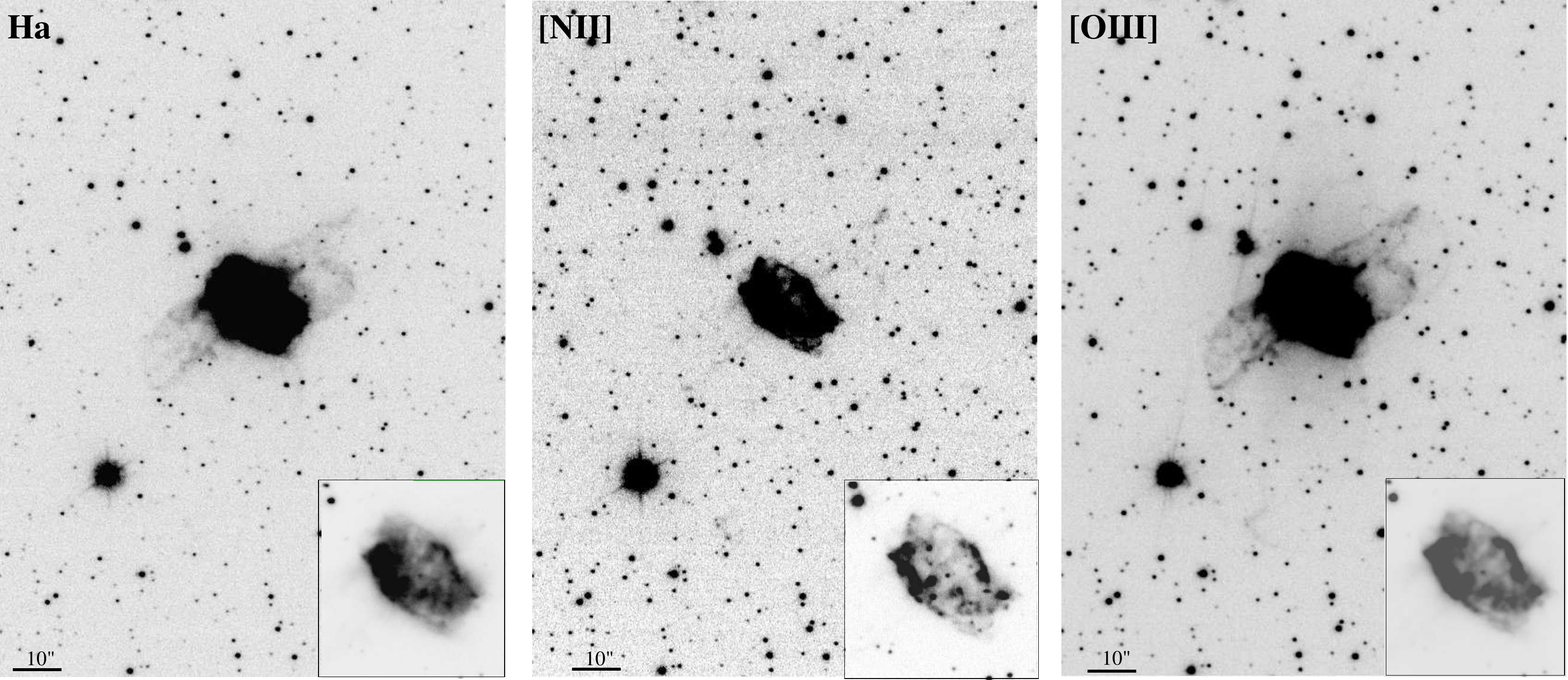}
\caption{
Detailed gray-scale rendering of the H$\alpha$, [N~{\sc ii}], and
[O~{\sc iii}] NOT images.
In the largest images (FoV $\simeq$1.7\arcmin$\times$2.2\arcmin) the
gray-scale levels are chosen to unveil the faintest and most diffused
structures of the PN, while saturating the central regions.
A number of outflows and filaments are seen mostly in [O~{\sc iii}].
The inset on each figure shows a zoom of the elliptical central region that
underlines its inhomogeneity with various high density clumps best seen in
[N~{\sc ii}] distributed around the central star.
}
\label{HaN2O3}
\end{center}
\end{figure*} 

\begin{figure*}
\begin{center}
\includegraphics[height=6cm]{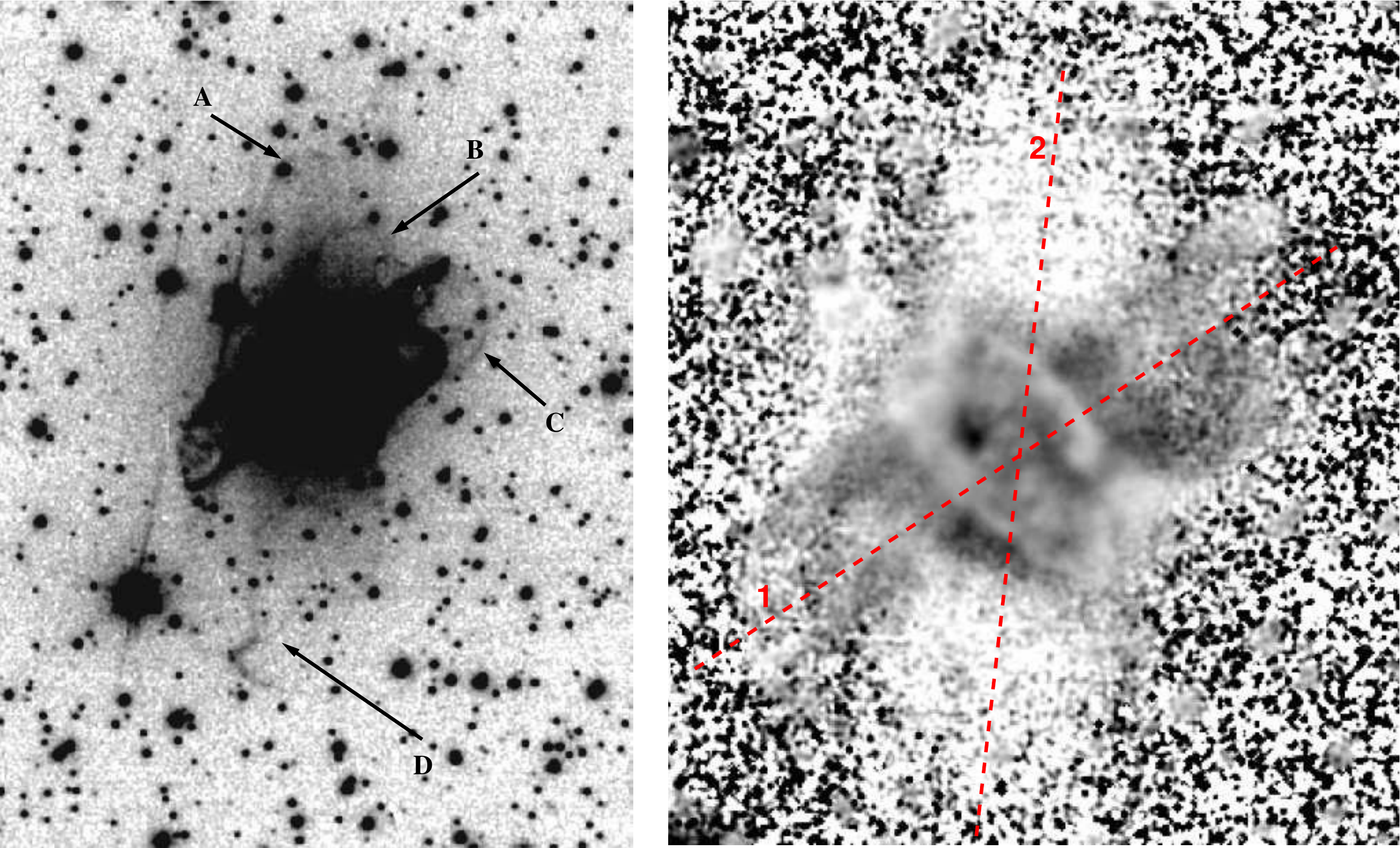}
\caption{
Left panel:
[O~{\sc iii}] image (FoV $\simeq$1.7\arcmin$\times$2.2\arcmin) with grey-scale
levels chosen to highlight the faintest patterns such as the arcs or outflows
tips labelled A, B, and C and the filamentary structure D.
Right panel: Gaussian filtered ratio map 
(H$\alpha$+[N~{\sc ii}])/[O~{\sc iii}]
(FoV $\simeq$1\arcmin$\times$1.2\arcmin) where the lighter zones imply
dominant [O~{\sc iii}] emission over H$\alpha$+[N~{\sc ii}].
The two dotted lines indicate the two main regions or axis showing
different ionization degrees.
}
\label{Faint_structures}
\end{center}
\end{figure*}

\begin{figure}
\begin{center}
\includegraphics[width=6cm]{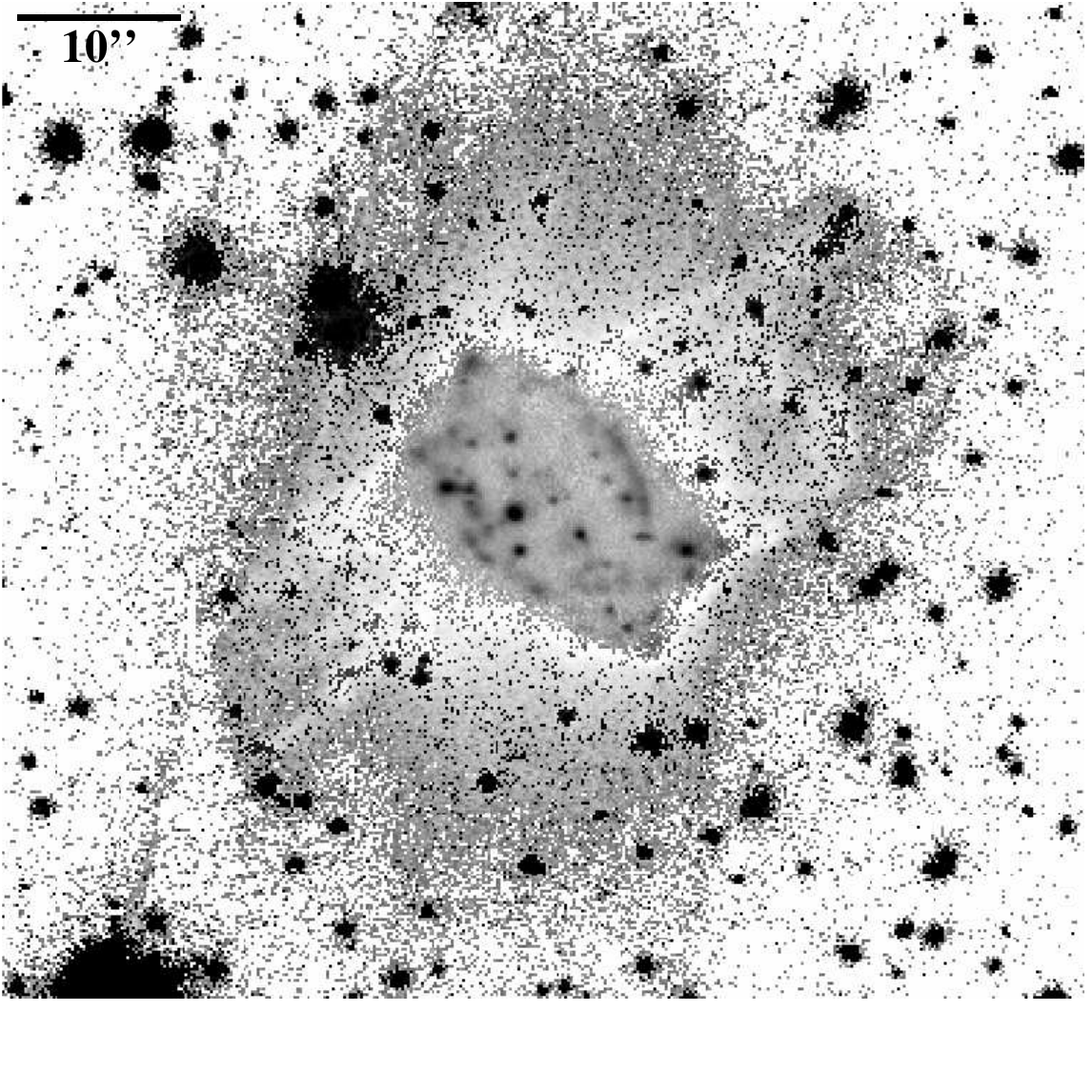}
\vspace{-0.5cm}
\caption{
[N\,{\sc ii}]/[O\,{\sc iii}] ratio map underlying the bright knots and
clumpy medium in the central part of PC\,22.
}
\label{ratio}
\end{center}
\end{figure}

Direct narrow-band images of PC\,22 were obtained on 2009 September 4
using the Alhambra Faint Object Spectrograph and Camera (ALFOSC) at
the 2.5m Nordic Optical Telescope (NOT) at the Roque de los Muchachos
Observatory (ORM, La Palma, Spain). 
The detector was a 2048$\times$2048 EEV CCD (13.5\,$\mu$m pixel size)
with a plate scale of 0$\farcs$184 pix$^{-1}$ providing a field of view
of 6$\farcm$3$\times$6$\farcm$3. 
The H$\alpha$ (\mbox{$\lambda_c = 6567$ \AA}, \mbox{FWHM $= 8$ \AA )},
[N\,{\sc ii}] (\mbox{$\lambda_c = 6588$ \AA}, \mbox{FWHM $= 9$ \AA)}, and
[O\,{\sc iii}] (\mbox{$\lambda_c = 5007$ \AA}, \mbox{FWHM $= 30$ \AA)}
filters were used to acquire two 900\,s images in each filter.
The images were processed using standard {\sc IRAF} routines.
A mean spatial resolution of 0$\farcs$6 was achieved, as determined
from the FWHM of stars in the field.

\subsection{MES long-slit high resolution optical spectroscopy} 

Long-slit high dispersion optical spectra of PC\,22 were acquired with the
Manchester Echelle Spectrograph (MES, \citealt{Meaburn2003}) mounted on the
2.12m telescope at the Observatorio Astron\'omico Nacional, San Pedro
M\'artir (OAN-SPM, Mexico).
A first set of data was obtained on 2015 August 14-16.
At this time, a 2048$\times$2048 pixels E2V CCD with a pixel size of
13.5 $\mu$m\,pixel$^{-1}$ was used as detector.
The fixed slit length is 6$\farcm$5 and we set the slit width to 150 $\mu$m
(1$\farcs$9).
We used 2$\times$2 and 4$\times$4 binnings leading to spatial scales of
0$\farcs$351\,pixel$^{-1}$ and 0$\farcs$702 \,pixel$^{-1}$, respectively.
The different spectra were taken with exposures of 1200~s and 1800~s
through two filters: an H$\alpha$ filter with $\Delta\lambda$ = 90 \AA\
to isolate the 87th order (0.05 and 0.1 \AA\,pixel$^{-1}$ spectral
scale for the 2$\times$2 binning and 4$\times$4 binning, respectively)
and an [O\,{\sc iii}]5007 \AA\ filter with $\Delta\lambda$ = 50 \AA\ to
isolate the 114th order (0.043 and 0.086 \AA\,pixel$^{-1}$ spectral scale
for the 2$\times$2 binning and 4$\times$4 binning, respectively).
We mostly used the latter filter based on the dominant [O\,{\sc iii}]
emission in the PN (as seen in the images, {\it see below}). Additional spectra were taken on 2016 February 25 using the same settings.

All the spectra were bias and flat corrected, and wavelength calibrated
using standard {\sc IRAF} routines for long-slit spectroscopy.
The wavelength-calibration was performed with a ThAr arc lamp to an
accuracy of $\pm$1km~s$^{-1}$.
The FWHM of the arc lamp emission lines, $\simeq$12$\pm$1km~s$^{-1}$,
provides an estimate of the spectral resolution.

\subsection{ESPRESSO multi-order optical echelle spectroscopy}

Multi-order echelle spectroscopic observations were performed on 2014 May
29 using ESPRESSO, an Echelle spectrograph mounted on the 2.12m telescope
at the OAN-SPM (Mexico).
The E2V CCD detector described above was used.
The wavelength range covered was 3900--7290 \AA\, distributed through
27 orders with a spectral scale $\sim$0.1 \AA\,pixel$^{-1}$ at 5000
\AA.
The slit was positioned at an angle of 47\degr\ passing through the
central star. No binning was applied and the exposure time was set
to 1800~s.
The data reduction, which includes bias removal and flat fielding
correction, was performed with {\sc MIDAS} \citep{Grosbol1989}.  
Two-dimensional spectra of each order was then extracted and wavelength
calibrated with a ThAr arc lamp using {\sc IRAF} routines.
We note the absolute wavelength calibration of the spectral orders
covering the [O~{\sc iii}] $\lambda$5007 and [Ar~{\sc v}] $\lambda$7006
emission lines was not optimal.  
No flux calibration was performed.
Each order was then analyzed separately and no attempt to bring
them together was made.

\begin{figure*}
\begin{center}
\includegraphics[width=8.5cm]{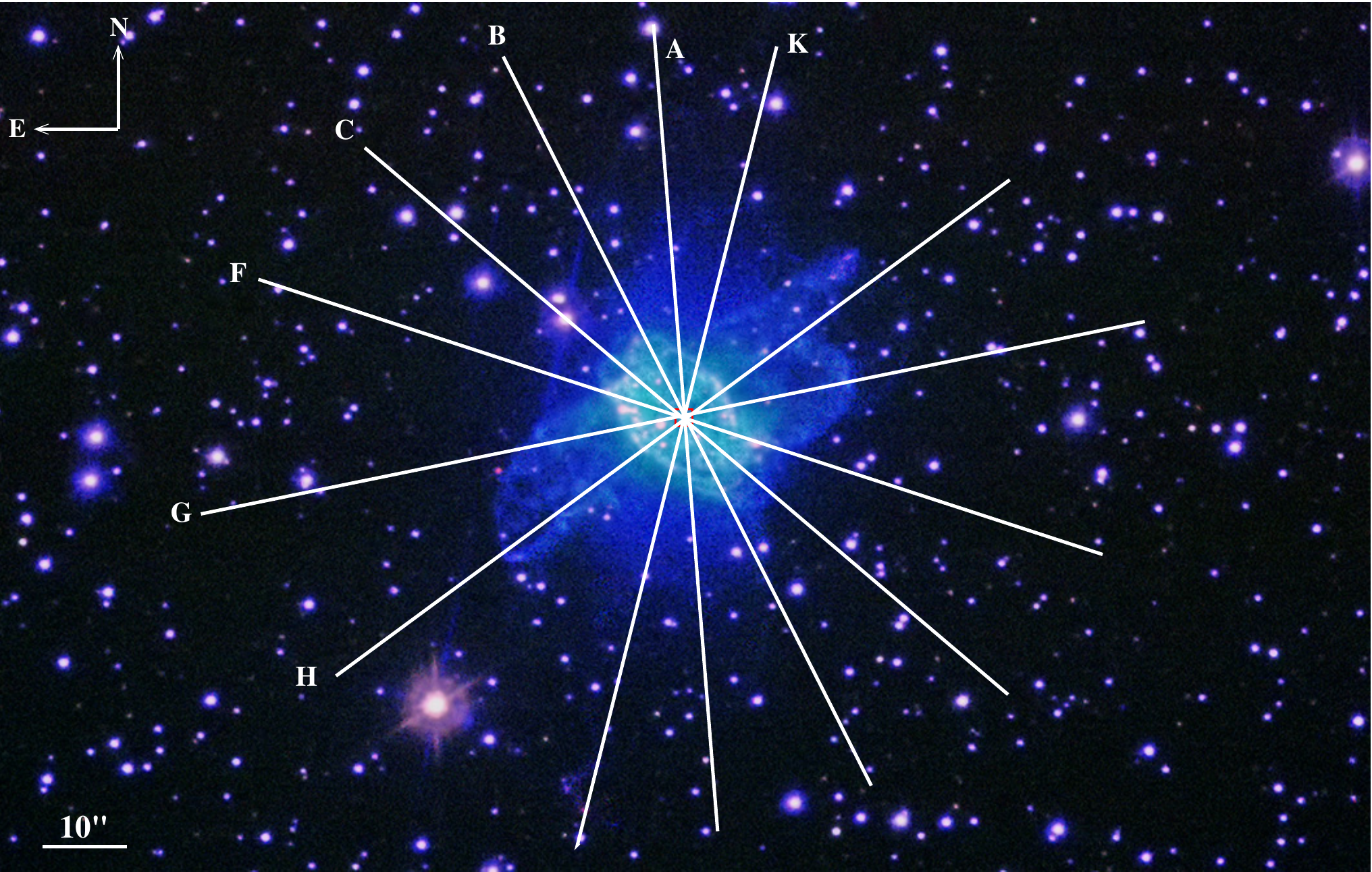}
\includegraphics[width=8.5cm]{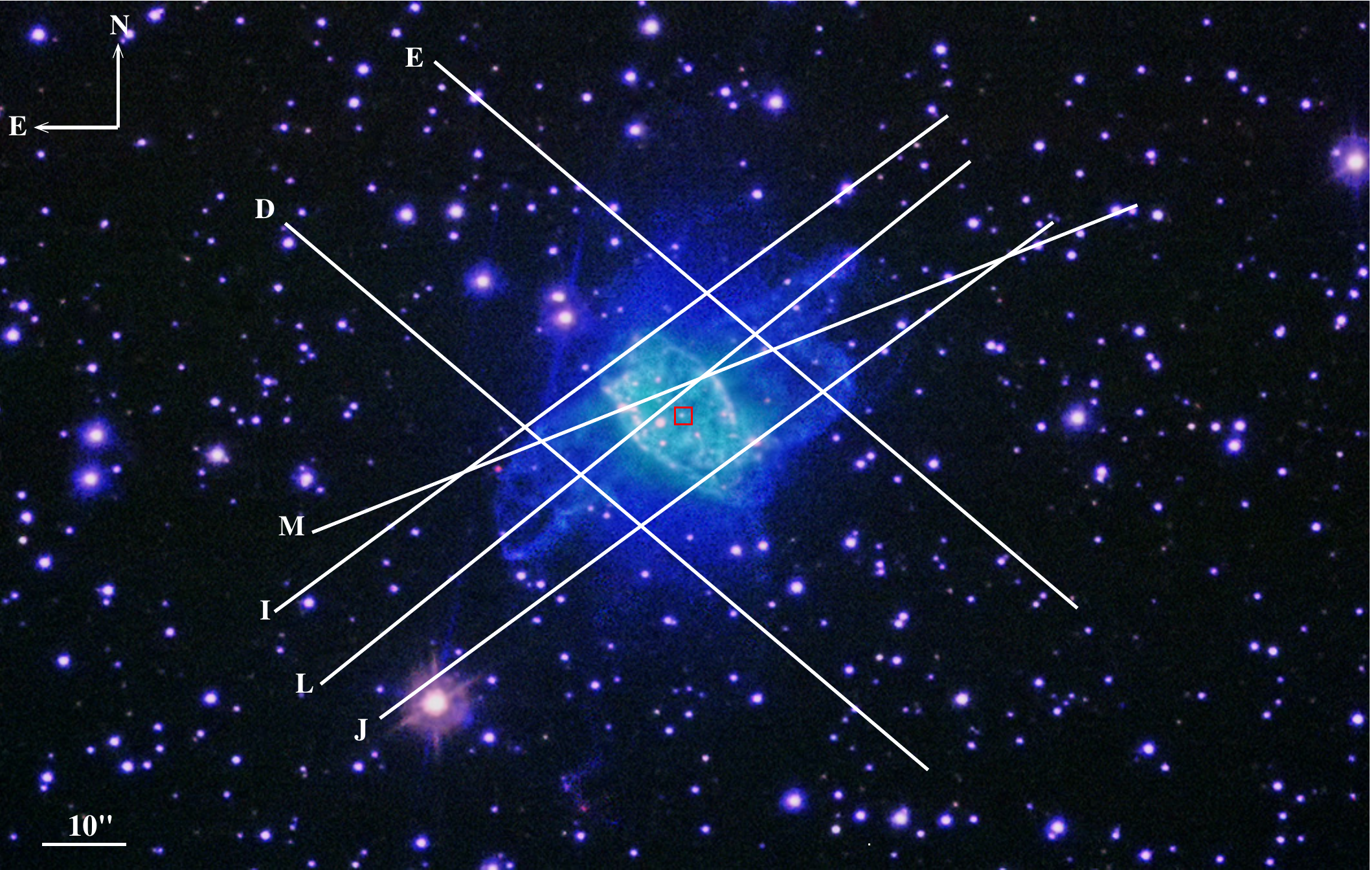}
\caption{
Slit positions used for the MES observations superimposed on the optical
RGB picture of PC\,22.
On the left panel we present all those passing through the central star (red square) and the remaining are shown on the right panel. The thirteen positions are meant to provide the best coverage of the PN for the kinematic analysis. }
\label{slits}
\end{center}
\end{figure*}

\section{Results}\label{sec_res} 
\subsection{Morphology}\label{res_morpho}

Our deep and high resolution H$\alpha$, [N\,{\sc ii}] and [O\,{\sc iii}]
images of PC\,22 unveil a detailed picture of a complex multipolar morphology.
The ``elliptical'' structure once believed to form the whole nebula
is actually the central part of a PN with multiple outflows 
(Fig.~\ref{RGB}).
We identified six large lobes or outflows (numbered 1-6 in Fig.~\ref{RGB})
and three secondary lobes or outflows closer to the central elliptical
region (numbered 7-9 in Fig.~\ref{RGB}). 
The [O\,{\sc iii}] 5007\AA\ nebular emission is significantly bright and shows an extension of $\sim$1$\farcm$5$\times$0$\farcm$9 including the tip of the faintest components presented in Fig.~\ref{HaN2O3} and Fig.~\ref{Faint_structures}.
The high level of [O\,{\sc iii}] 5007\AA\ emission is somewhat peculiar as it extends up to the outermost regions of the PN and is not confined to the inner parts of the nebula as one would expect (i.e. there is no stratification in the emission lines distribution).
Such behavior, however, is not unique and it has been noted in other PNe
such as NGC\,6309 \citep{Rubio2015} and NGC\,6905 \citep{Cuesta1993}.

The faint aforementioned structures are under the form of filaments
(in the south for example) and weak outflows which are best seen in
the contrast-enhanced image shown in Fig.~\ref{Faint_structures}.
Three arc-like patterns, labelled ``A, B and C'' can be associated to the
tips of three different outflows detailed below.

The H$\alpha$ and [O\,{\sc iii}] images indicate a bright and slightly
collimated group of ejecta along the SE-NW direction (hereafter Axis 1),
with the NW outflow appearing to be ``broken'' by the emergence of a new
one whose much dimer rim is mostly seen in [O\,{\sc iii}]
(Fig.~\ref{Faint_structures}-- label ``C'').
This effect is not symmetric, i.e.\ the structure ``C'' does not have a
clear counterpart in the SE outflow.
In the N-S direction (hereafter Axis 2), the presence of a bright inner
section immediately followed by a much fainter one at larger distances
(seen in [O\,{\sc iii}]) could indicate the occurrence of two different
outflows running approximately along the same direction.
The presence of these new structures is supported by the arc-like patterns
``A'' and ``B'' observed in Fig.~\ref{Faint_structures}.
Once again, there is not a clear, symmetric counterpart to the northern
outflows.   
Fig.~\ref{Faint_structures} also shows that the ejecta in the Axis 2
are wider, less collimated, less defined and generally fainter than
those in Axis 1.
There is also a slight difference in the dominating emission lines (and
hence in the ionization factor).
Whereas [O\,{\sc iii}] emission prevails in Axis 2, a combination of
H$\alpha$ and [N\,{\sc ii}] emissions are prevalent in Axis 1.
A more exhaustive analysis of these two regions, from a chemical point
of view, will be presented in the second part of this investigation
(Sabin et al.\ in preparation).

The central region of PC\,22 has a size of 13$\farcs$5$\times$23$\farcs$9
and a position angle (P.A.) on the plane of the sky of 49\degr.
It shows a notable inhomogeneous morphology in all emission lines
(Fig.~\ref{HaN2O3}).
The ``clumpy''/knotty nature of this region is better seen in [N\,{\sc ii}],
which is a known tracer of low ionization structures compared to the higher
ionization [O\,{\sc iii}] emission line \citep{Corradi1996}.
We produced different line ratio images, namely (H$\alpha$+[N\,{\sc ii}])/[O\,{\sc iii}] and [N\,{\sc ii}]/[O\,{\sc iii}], to check for the presence of such patterns (Fig.~\ref{ratio}).
No new structures were revealed besides those already seen.
In terms of spatial distribution, the clumps described above are seen
inside and (mostly) at the edges of the elliptical ring, i.e., the
inhomogeneities are concentrated in the central region.
More precisely the main over-density positions are coincident with the base
of the ejecta related to the Axis 1, suggesting the irregular brightening or
ionization of material following the passage of a collimated outflows.
Such interaction is not unusual and has been described by \citet{Vaz2000}
in M\,2-48. 

The morphological analysis of PC\,22 unveils a much more complex object
than initially described by an elliptical morphology.
The PN shows multiple non symmetric ejecta with two main differing degrees
of ionization whose ``waist'' seems to be shaped by the interaction with
the different outflows.

\subsection{Kinematics }

\begin{table}
\begin{center}
\caption{
\label{characteristics}
Description of the slits used for the MES observations.
''Offset CSPN'' corresponds to the slit offset from the central star.
}
\begin{tabular}{@{\extracolsep{4pt}}|l|c|l|c|c|}
\hline
Slit & P.A.    & Filter & Offset CSPN & T$_{exp}$  \\
     & (\degr) &        &  (\arcsec)  &  (s) \\
\hline
A    &    5     &  [O\,{\sc iii}] &  0  &  1800     \\ 
B    &    28    &  [O\,{\sc iii}] &  0  &  1800      \\
C    &    49    &  [O\,{\sc iii}], H$\alpha$& 0  &  1200  \\
D    &    49    &  [O\,{\sc iii}] & 13.3   &  1200   \\
E    &    49    &  [O\,{\sc iii}] & 13.0   &  1200   \\
F    &    75    &  [O\,{\sc iii}] & 0   &  1200    \\    
G    &    100   &  [O\,{\sc iii}] & 0   &  1200   \\   
H    &    137   &  [O\,{\sc iii}], H$\alpha$  & 0 &  1200 \\   
I    &    137   &  [O\,{\sc iii}] & 10.0 &  1200   \\
J    &    137   &  [O\,{\sc iii}] & 8.0 &  1200   \\
K    &    166   &  [O\,{\sc iii}] & 0 &  1200 \\
L    &    128   &  [O\,{\sc iii}] & 2.1 &  1800 \\
M    &    114   &  [O\,{\sc iii}] & 3.4 &  1800 \\
\hline
\end{tabular}
\end{center}
\end{table}

\begin{table}
\caption{
Kinematical characteristics (i.e.\ velocities and age) for eight different
structural components obtained from our best fitting model constructed with
{\sc shape} (see Fig.\ref{main_structures2}).
Velocities are relative to the calculated systemic of 36.4 km~s$^{-1}$
and $\theta$ is the angular radius measured from the central star.
We note that the de-projected velocity (V$_{dep}$) for structure 7 is
likely to be ``affected`` by the presence of other velocity field
components from its surrounding.
}
\begin{tabular}{|c|r|r|c|c|c|}
\hline
\multicolumn{1}{|l|}{Structures} &
\multicolumn{1}{c|}{V$_{rad}$}  &
\multicolumn{1}{c|}{$\theta_{dep}$}  &
\multicolumn{1}{c|}{{\it i}} &
\multicolumn{1}{l|}{V$_{dep}$} &
\multicolumn{1}{c|}{t$_{kin}$} \\ 
\multicolumn{1}{|c|}{number} &
\multicolumn{1}{l|}{(km~s$^{-1}$)} &
\multicolumn{1}{c|}{(\arcsec)} &
\multicolumn{1}{c|}{(\degr)}  &
\multicolumn{1}{l|}{(km~s$^{-1}$)} &
\multicolumn{1}{c|}{(yrs)}  \\ \hline
1 &  --9$\pm$3~~~ & 12.1~ &  83 & 131~~~~~ & 2300$\pm$700~ \\ 
2 & --40$\pm$7~~~ & 13.2~ &  74 & 140~~~~~ & 2400$\pm$800~ \\ 
3 &  --9$\pm$4~~~ &  8.6~ & 100 &  56~~~~~ & 3800$\pm$1400 \\ 
4 &   21$\pm$4~~~ &  8.3~ & 100 &  59~~~~~ & 3500$\pm$1300 \\ 
5 &   23$\pm$3~~~ & 15.1~ & 103 & 101~~~~~ & 3700$\pm$1200 \\ 
6 &   26$\pm$5~~~ & 23.0~ & 100 & 155~~~~~ & 3700$\pm$1100 \\ 
7 &   12$\pm$3~~~ &  5.4~ & 100 &  33:~~~~ & 4100$\pm$2000 \\ 
8 &  --7$\pm$3~~~ &  6.5~ & 100 &  48~~~~~ & 3400$\pm$1400 \\
\hline
\end{tabular}
\label{Velocities_full}
\end{table}
\subsubsection{Morpho-kinematical modelling of MES data with {\sc SHAPE}}\label{res_MES}

The large set of slit positions (see Fig.~\ref{slits} and
Tab.~\ref{characteristics}) provides a unique coverage of PC\,22
to carry out a comprehensive kinematical analysis.
The observations used in the following will be those taken in the
[O\,{\sc iii}] line.
Using the seven positions passing through the central star, we derived a
mean observed systemic velocity $\bar{V}_{obs}$ of 28.1 km\,s$^{-1}$, which
corresponds to an LSR mean value of 36.4 km\,s$^{-1}$.

\begin{figure}
\begin{center}
\includegraphics[width=8cm]{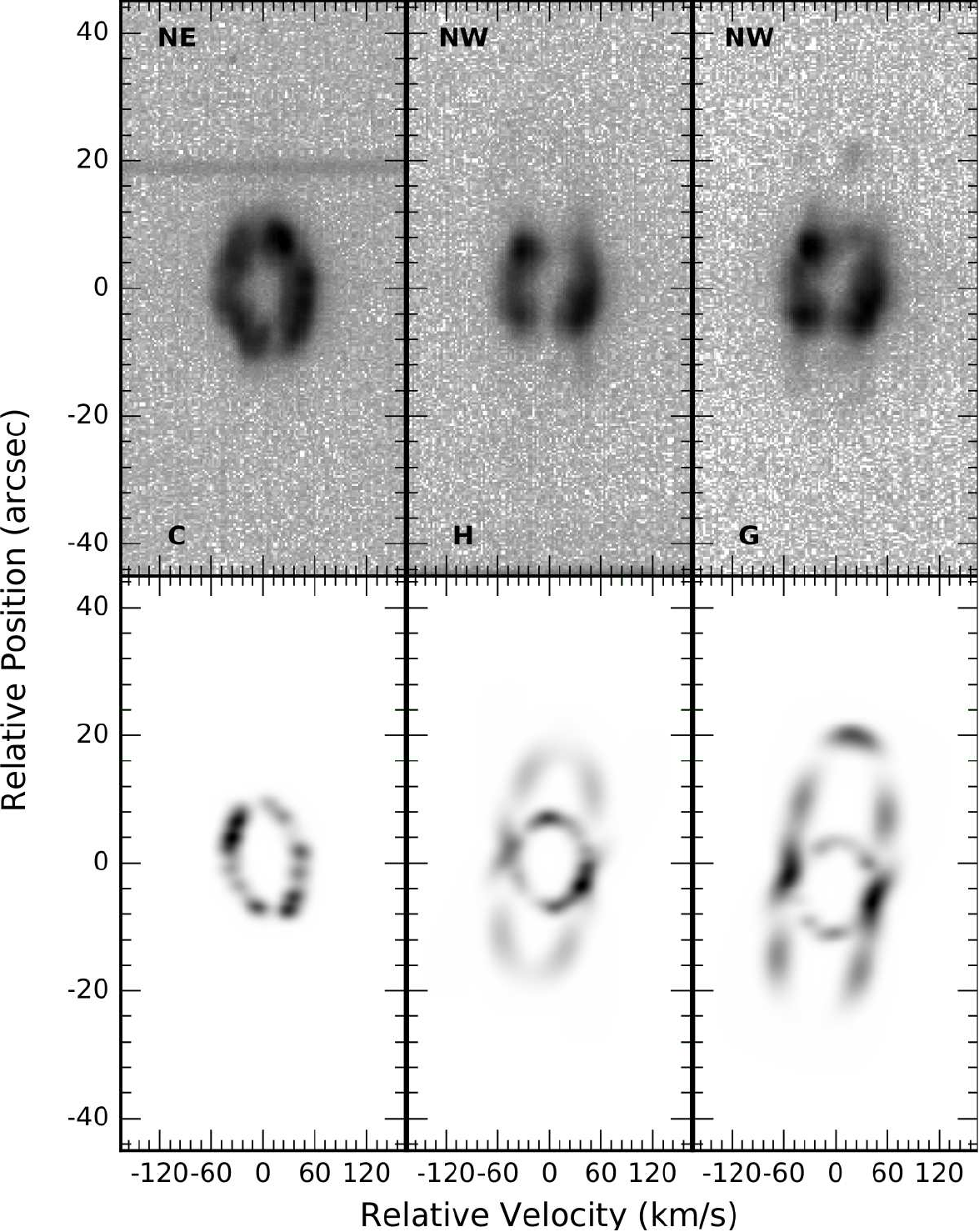}
\caption{
Top: MES [O\,{\sc iii}] position-velocity (PV) plots of three of the
main slit positions with higher signal-to-noise level, namely slit
positions C, G, and H (see Tab.~\ref{characteristics} and Fig.~\ref{slits},
where the C and H slits go across the nebula center along P.A.\ 49\degr\
and 137\degr, respectively, and slit H is offset form the central star and
goes along P.A.\ 100\degr.
The binning factor is 2$\times$2 in all images.
Bottom:
{\sc Shape} kinematical modelling of these lines for the model assumed in the
text.
}
\label{main_structures1}
\end{center}
\end{figure}

Although we were able to secure a good mapping of the PN, the overall higher brightness of the central region and the lobes corresponding to the Axis 1 were the only structures to actually appear clearly in the spectra, with the other components being much fainter.
The upper panels of Fig.~\ref{main_structures1} show the resulting spectra or position-velocity (PV) maps, passing through the short and long axis of the main internal structure (PA=49\degr: slit C and PA=137\degr: slit H, respectively) and the bright lobes in Axis 1 at PA=100\degr\ corresponding to the slit G.
In all three cases we note the irregular shape of the inner regions which
corresponds to the presence of compact knots.
The remaining observed PV maps are presented in the appendix A, Fig. \ref{app1} and \ref{app2}.

\begin{figure}
\begin{center}
\includegraphics[width=8cm]{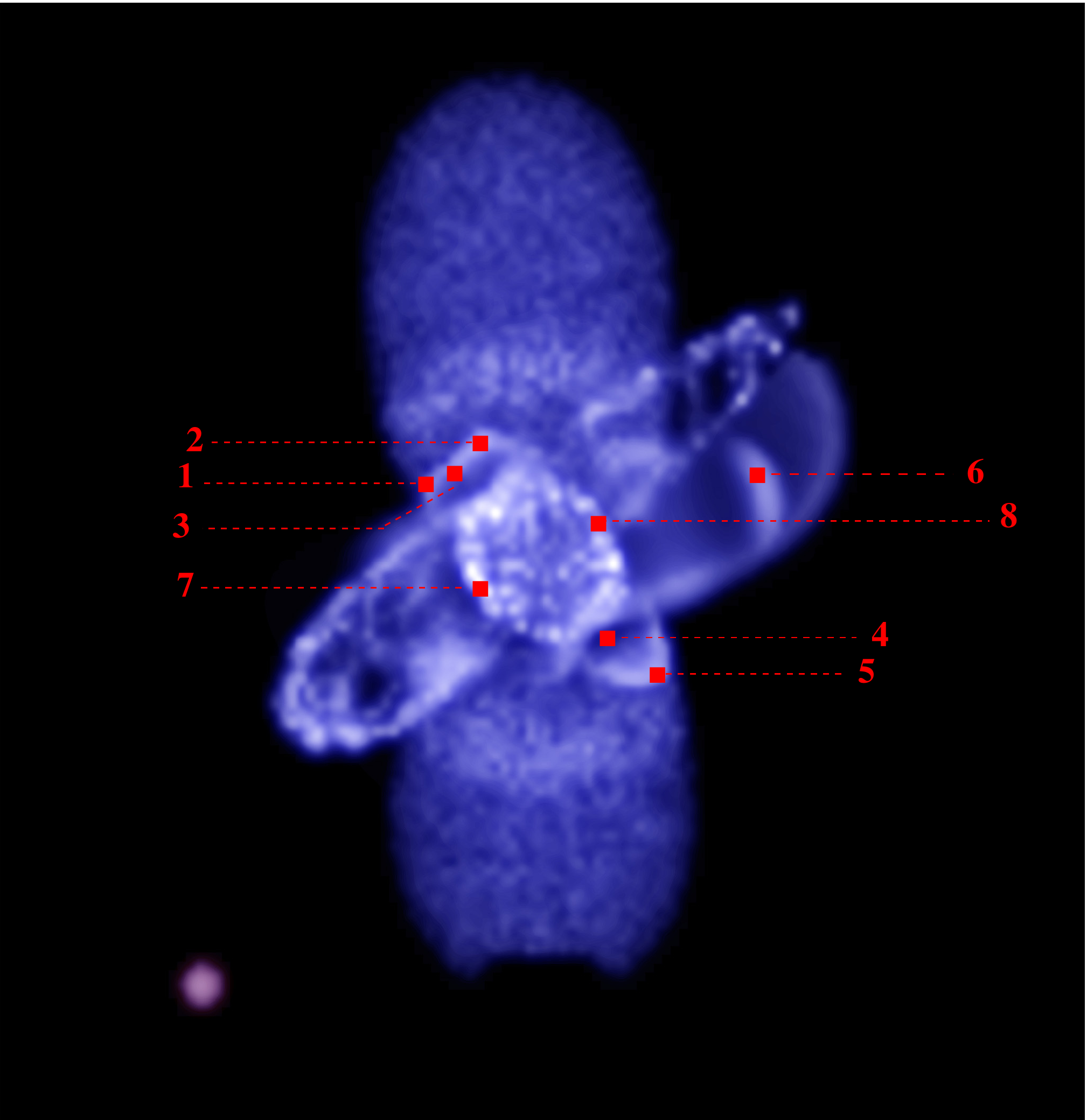}
\caption{
{\sc Shape} reconstruction of PC\,22 based on all PV diagrams.
The numbers indicate the structures of reference for the different
kinematical calculations (see table \ref{Velocities_full}).
}
\label{main_structures2}
\end{center}
\end{figure}

We subsequently used the morpho-kinematical tool {\sc Shape}
\citep{Steffen2011} to model all the PV diagrams and obtain
a complete view of the (kinematical and morphological) structure
of PC\,22.
Due to the inhomogeneous nature of the central part, we concentrated only
on the representation and modelling of the basic velocity patterns
corresponding to an equatorial ring and several bipolar outflows.  
Therefore, one will observe a difference between the actual (thicker)
observed spectra and the narrower synthetic ones (see the lower panels
of Fig.~\ref{main_structures1}).  
The non-inclusion of the knots/clumps does not hamper the overall kinematical analysis. We should keep in mind that the spectra are not treated independently i.e. a change in the position/morphology or the velocity of one spectrum will affect the position/morphology or the velocity of the adjacent one(s).
The full set of PV diagrams are shown in Fig.~\ref{main_structures1} and the appendices \ref{app1} and \ref{app2}.
The subsequent {\sc Shape} reconstruction of the PN is shown in
Fig.~\ref{main_structures2}.
Table~\ref{Velocities_full} summarizes the data for the brightest (and best defined) components. In this case we derived information for the two northern ``bumps'' (\#1,\#2), the southern small outflow (\#5), the four extensions of the internal ellipsoid (\#3,4,7 and 8) and finally one of the western outflows (\#6).
In all cases the velocities were measured at the tip of the structures. For each we calculated the radial velocities (V$_{rad}$) and their respective errors, the de-projected angular radii ($\theta_{dep}$), the inclinations ({\it i}), de-projected velocities (V$_{dep}$) and kinematical ages (t$_{kin}$). The errors on {\it i} and V$_{dep}$ were derived empirically solely based on our model and we estimated a mean error on the inclination of $\sim$5\degr\, and  a mean error on V$_{dep}$ of $\sim$3 km\,s$^{-1}$. Those are the typical values above which we observed a notable variation in the PV maps. The inclinations found indicate that most of the structures are nearly lying in the plane of the sky. The deduced de-projected velocities allow us to distinguish two groups: the medium velocities structures, from  $\sim$33 to 59 km\,s$^{-1}$ (although we note that the former value is very likely affected by nearby structures), which are related to the central region, and the high velocities (jets-like) ones, above $\sim$100 km\,s$^{-1}$ related to the ``bumps'' and larger outflows.

The kinematical age of these components can be derived using the
relationship: 
\begin{equation}
\hspace{2cm} t_{kin}= \frac{4744 * d (kpc) * \theta (\arcsec)}{V_{dep} (km/s)}.
\end{equation}
We can clearly see two groups: 
components \#1 and \#2, related to the northern ``bumps'' or small
outflows, have a mean age of 2300$\pm$800 yr, whereas the other
components have a mean age of 3600$\pm$1300 yr\footnote{
Component \#7, which suffers from the presence of adjacent knots that
add velocity components not necessarily related to this part of the
shell, has not been taken into account in this average.
}.The elliptical region, if interpreted as a ring, has a kinematical age of $\sim$4500 yrs. Within the errors, the different components are coeval, but we notice
that the interaction of the outflows with the circumstellar medium would reduce their
expansion velocity (thus increasing their kinematical age).
The kinematical analysis completes the morphological one, pointing towards
a PN mostly shaped by the interaction of high-velocity non-symmetric winds,
the zone of the interaction being the central region which as a consequence
is highly inhomogeneous.
Most of the events (e.g. outflows launching) also appear to have occurred
at approximately the same time in the life of the not so old PC\,22.
This will give us an important constraint on the evolutionary history of this PN {\it (see section \ref{sec_dis})}.

\subsubsection{ESPRESSO}\label{res_REOSC}

\begin{figure*}
\begin{center}
\advance\leftskip-1cm
\includegraphics[width=20cm]{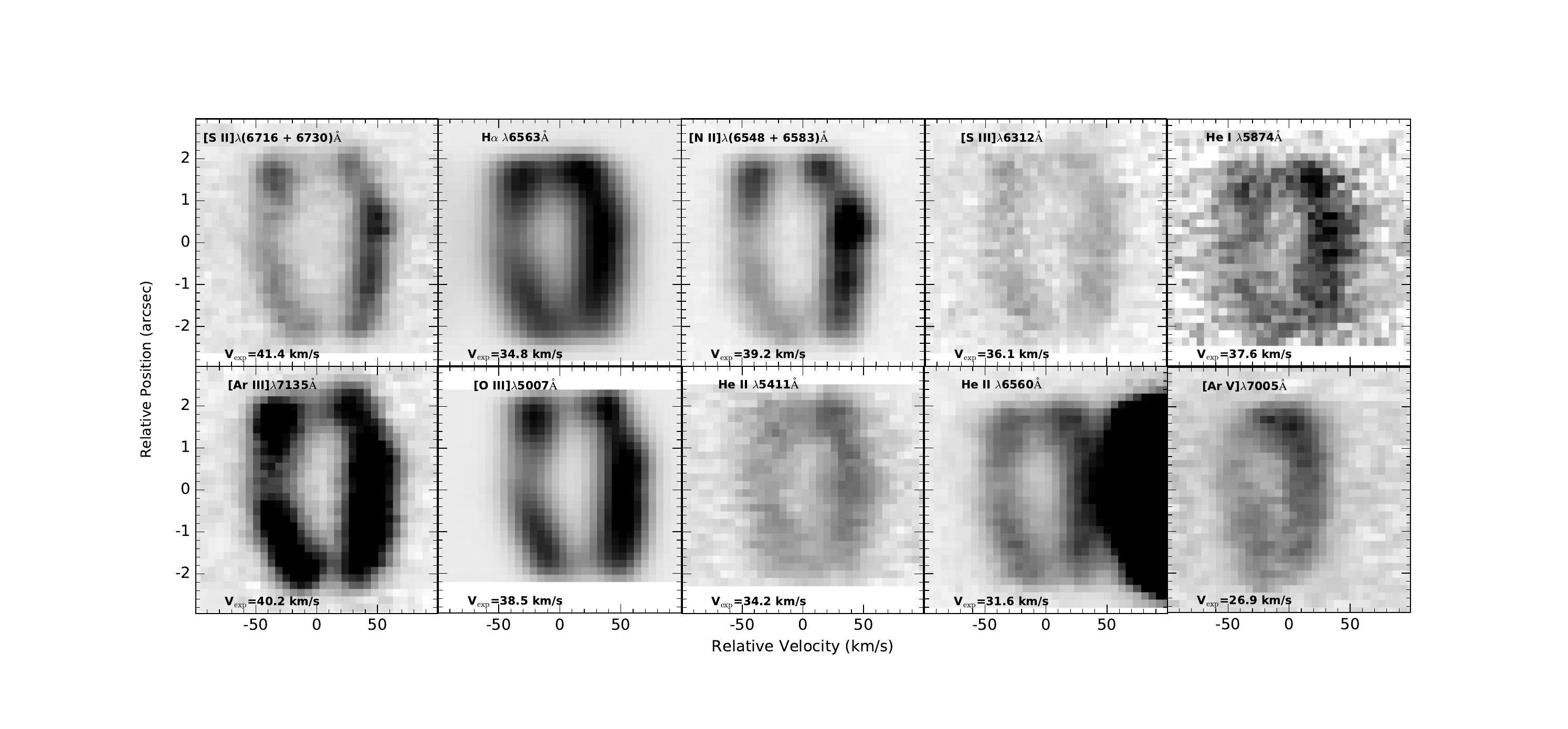}
\vspace{-1cm}
\caption{
Position-Velocity (PV) diagrams resulting of the analysis of twelve well 
defined emission lines from the ESPRESSO echelle spectra of PC\,22.
All lines show a split and the corresponding expansion velocities are quoted.
The PVs are arranged in terms of increasing potential ionization level so
as to visually follow the change in V$_{exp}$.  
We notice that the [S\,{\sc ii}] and [N\,{\sc ii}] emission lines have
been added to increase the signal-to-noise ratio.
As for the He II (6560 \AA) line, the dark area on the right hand-side
corresponds to the much brighter contiguous H$\alpha$ line.
}
\label{vel_gauss1}
\end{center}
\end{figure*}

The inspection of the ESPRESSO echelle spectrum of the central region of
PC\,22 reveals a notable number of emission lines that present blue and
red components (Fig.~\ref{vel_gauss1}).
These can be interpreted as the approaching and receding walls of the
central structural component, respectively, and therefore they provide
information on its internal kinematics.
A list of these lines is reported in Table~\ref{reosc_vel}.
The radial velocities associated with each line has been determined by
fitting Gaussian curves to each component to derive their wavelengths.
We estimated that the maximum error on the line measurement is
$\sim$6 km\,s$^{-1}$.
Table~\ref{reosc_vel} reports the observed ($V_{\rm obs}$), LSR corrected ($V_{\rm LSR}$),
and relative (from systemic velocity, $V_{\rm rel}$) velocities for both
blueshifted and redshifted components.
The systemic velocity, as derived from the ESPRESSO data, is 36.4 km\,s$^{-1}$,
which is consistent with that derived from the MES data.

\begin{figure}
\begin{center}
\includegraphics[width=7cm]{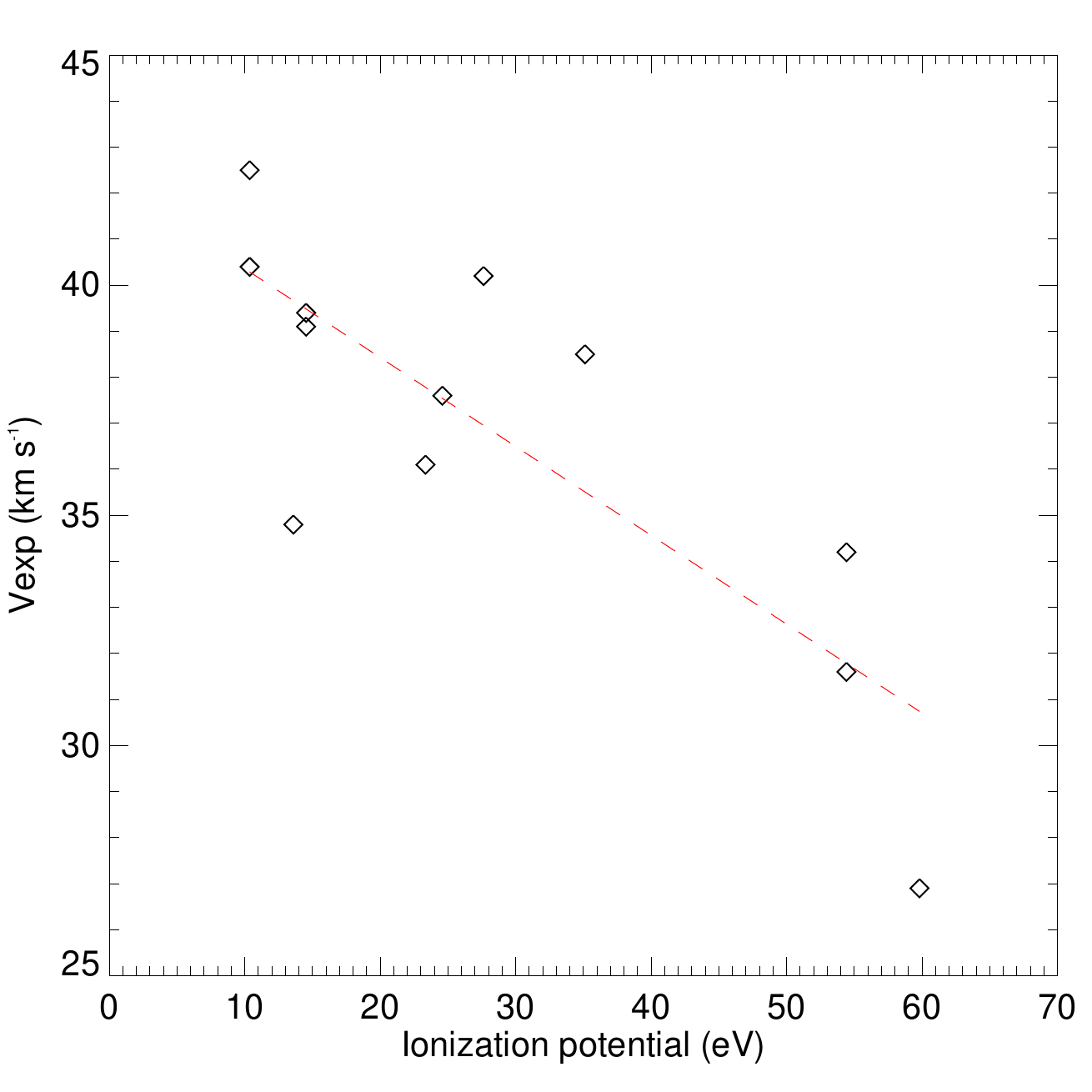}
\caption{Graph render of Table \ref{wilson} with the red dotted line representing the linear fit.}
\label{Wilson_graph}
\end{center}
\end{figure}

The expansion velocities are different for different lines.
We note that there is a trend for the expansion velocity to increase
as the ionization potential of the species increases, as
revealed by the linear fit to the data (Table~\ref{wilson} and
Fig.~\ref{Wilson_graph}).
Since species with higher ionization potential are closer to the CSPN, it
means there is a negative velocity gradient as material is located further
away from the CSPN.
This is the so-called ``Wilson effect'' \citep{Wilson1950}, which describes
the decrease of the line separation (or expansion velocity) with an
increase of the ionization potential in PNe.

\begin{table}
  \caption{
    Kinematical analysis based on the different line splits observed in the ESPRESSO Echelle spectra at PA=47\degr.
    The values were derived through a Gaussian fit of the lines.
    For both the blue- and redshifted lines we present: the observed velocities (V$_{obs}$), the observed data corrected from LSR (V$_{LSR}$) and the velocities relative to the systemic (V$_{rel}$).
    Note that, due to difficulties in the absolute wavelength calibration of the spectral orders covering the [O~{\sc iii}] and [Ar~{\sc v}] emission lines, only the observed velocities are quoted for these lines.  
  }
\begin{tabular}{@{\extracolsep{4pt}}|l|c|c|c|c|c|c|}
\hline

\multicolumn{1}{|l|}{Line ID}  & \multicolumn{3}{c}{Blueshifted lines (km/s) } & \multicolumn{3}{c|}{Redshifted lines (km/s)}    \\  
\cline{2-4} \cline{5-7}
\multicolumn{1}{|l|}{} & \multicolumn{1}{c|}{V$_{obs}$} & \multicolumn{1}{c|}{V$_{LSR}$} & \multicolumn{1}{c|}{V$_{rel}$} & \multicolumn{1}{l|}{V$_{obs}$} & \multicolumn{1}{l|}{V$_{LSR}$} & \multicolumn{1}{l|}{V$_{rel}$} \\ \hline

{[O\,{\sevensize III}]} 5007 & -21.9 &-- & -- & 55.2 & -- & --  \\ 

{He\,{\sevensize II}} 5411 & -30.7 &6.6  & -29.8 &37.7 &74.9  &38.6    \\ 

{He\,{\sevensize I}} 5876 & -43.8 & -6.5 & -42.9 &31.4  & 68.7 &32.3    \\ 

{[S\,{\sevensize III}]} 6312 & -33.2 &4.1 & -32.3 &38.9  &76.2  & 39.8   \\ 

{[N\,{\sevensize II}]} 6548 & -37.2 & 0.1 & -36.3 & 40.9 & 78.3 & 41.8 \\ 

{He\,{\sevensize II}} 6560 & -34.0 & 3.3&-33.1  & 29.3 & 66.6 & 30.2   \\

H$\alpha$ 6563 & -39.4 &-2.1 &-38.5 &30.2  & 67.5 &31.1    \\

{[N\,{\sevensize II}]} 6583 & -44.8 &-7.5  &-43.9  & 33.9 & 71.3 & 34.9   \\ 

{[S\,{\sevensize II}]} 6716 & -42.1 & -4.8 & -41.2 &42.9 & 80.3 & 43.9  \\ 

{[S\,{\sevensize II}]} 6731 &  -39.3& -1.9 &-38.4 &41.6  & 78.9 & 42.5   \\ 

{[Ar\,{\sevensize V}]} 7005 & -41.9 & -- &--  & 12.0 & -- & -- \\ 

{[Ar\,{\sevensize III}]} 7135 & -37.5& -0.2 & -36.6 &42.8  &80.2  & 43.8   \\
\hline
\end{tabular}
\label{reosc_vel}
\end{table}

\begin{table}
\begin{center}
\caption{Ionization potential {\it vs} expansion velocity for the line with clear splitting seen in PC\,22}
\begin{tabular}{|l|c|c|c|}
\hline
Ion &  Line  & Ionization potential  & V$_{exp}$\\
    & (\AA)  &(eV) &  (km\,s$^{-1}$) \\
\hline
{[S\,{\sevensize II}]}& 6716 & 10.36 & 42.5 \\
{[S\,{\sevensize II}]}& 6731 & 10.36 & 40.4 \\
H$\alpha$& 6563 & 13.59 & 34.8\\
{[N\,{\sevensize II}]}& 6548 & 14.53 & 39.1\\
{[N\,{\sevensize II}]} &6583 & 14.53 & 39.4\\
{[S\,{\sevensize III}]}& 6312 &    23.34   & 36.1 \\
{He\,{\sevensize I}}& 5876 & 24.59 &37.6 \\
{[Ar\,{\sevensize III}]}& 7135 & 27.63 & 40.2\\
{[O\,{\sevensize III}]}& 5007 & 35.12 & 38.5 \\
{He\,{\sevensize II}}& 5411 & 54.42 & 34.2 \\
{He\,{\sevensize II}}& 6560 & 54.42 & 31.6 \\
{[Ar\,{\sevensize V}]}& 7006 & 59.81 & 26.9 \\
\hline
\end{tabular}
\label{wilson}
\end{center}
\end{table}

\section{Main findings}\label{sec_find}

The main results of the morpho-kinematical study of PC\,22 presented here
are the following:\\
\begin{itemize}
\item
The PN, principally emitting in {[O~{\sc iii}]}, is not elliptical as
previously stated, but multipolar with several sets of non-symmetric
outflows showing different degrees of brightness, collimation and ionization.
\item
Most (if not all) over-density and low-ionization structures (e.g., the
[N~{\sc ii}] bright) are concentrated in the elliptical central part of
the nebula and are spatially related to the passage of the aforementioned
outflows.
\item
The morpho-kinematical model obtained with {\sc SHAPE} reveals information
on the different structural components of PC\,22.
Hence: \\
- The de-projected velocities of the outflows and bumps are rather large
($\geq$ 100 km\,s$^{-1}$) compared to that of the central elliptical
region ($\sim$30--60 km\,s$^{-1}$). \\
- All the structures share similar inclination around 100\degr\ (slightly
less for the structures \#1 and \#2 in Fig.\ref{main_structures2}), i.e.,
nearly all the ejections can be considered as coplanar and close to the
plane of the sky. \\
- The kinematical age calculations are also well consistent for all the
analysed outflows with a mean of 3600$\pm$1300 yrs, but for structures
\#1 and \#2, which seem to be slightly ``younger" (2300$\pm$800 yrs).
This may indicate either that structural features at the elliptical
central region have been accelerated, or that the outflows are been
slowed down (see, for instance, the ejecta in the born-again PNe
A\,30 and A\,78, \citealt{Fang2014}).
This is a tantalizing possibility, but at this moment we conclude that
all ejections in PC\,22 should be considered to be coeval, owing to the
uncertainties on the age calculation.
\item
The ESPRESSO echelle data obtained for twelve emission lines presenting
blue- and red-shifted components allowed us to derive their expansion
velocities.
This varies from 26.9 to 42.5 km\,s$^{-1}$ and tends to abide by the
``Wilson effect'' as we noted a decrease of the expansion velocity
with an increase of the ionization potential.
\end{itemize}

\section{Discussion and conclusion}\label{sec_dis}

The combination of the morphological and kinematical analysis indicates
that PC\,22 belongs to the category of so-called {\bf starfish PNe} at
a late evolutionary stage.
Starfish nebulae were introduced by \citet{Sahai2000} to describe Hen\,2-47
and M\,1-37, two young PNe investigated during an H$\alpha$ imaging survey
with the \emph{Hubble Space Telescope}.
Later, the proto-PN IRAS\,19024$+$0044 \citep{Sahai2005} was added to
this group.
All three objects present the same morphological characteristics: they show
multiple well defined and fast ($\sim$100--200 km\,s$^{-1}$) collimated lobes
with approximately the same sizes which would indicate that they have
emerged at the same time and an equatorial waist or ring.
Our claim that PC\,22 is very likely an evolved version of the
aforementioned nebulae holds in the fact that:
\begin{enumerate}
\item
The contemporary lobes have a larger extent ($\sim$20--40\arcsec), they
are fainter and show less collimation in some cases (see Axis 2).
Therefore, the compactness of the young starfish nebulae (with outflow
extents less than $\sim$5\arcsec\, in all cases) and their sharpness
of their lobes are lost.
\item
PC\,22 is a high-excitation PN, with dominant {[O~{\sc iii}]} emission,
indicating that the central star is hot and the photoionization very
efficient.
This is contrary to the proto-PNe and young PNe starfish, which were
selected for the \emph{HST} observing program HST-P6353 based on the
(very) low {[O\,{\sevensize III}]}5007 \AA\ to H$\alpha$ line ratio,
probing their ``early'' evolutionary stage.
PC\,22 would therefore join NGC 6058 \citep{Guillen2013} in the very
small group of evolved starfish PNe.
\end{enumerate}

This investigation opens a new door in the understanding of the history
formation and evolution of such objects.
The morphology of PC\,22 can be explained by a fast and probably primarily
uniform wind from the Asymptotic Giant Branch (AGB) phase going through
irregular and inhomogeneous cavities from the previous denser AGB shell.
\citet{Steffen2013} studied this phenomenon numerically using {\sc shape}
and an adjustable turbulent noise structure (Perlin noise) to simulate,
via hydrodynamical modelling, a filamentary AGB shell (their Figure 3)
within which a fast wind evolves.
By changing the size of the cells or voids of the filamentary sheet, they
could reproduce different types of starfish nebulae, indeed showing non
symmetric patterns.
It resulted, that the smaller the cells, the smaller the resulting outflows
and reversely with a loss of collimation (their Figure 4).
\citet{Steffen2013} also found that starfish nebulae would only occur if
the AGB shell and post-AGB wind keep a rather constant density and
velocity with respect to the angular distance from the equator.

PC\,22 would therefore be at a stage, still early in the $\sim$10$^{4}$ yrs
PN life, where the material dragged by the fast post-AGB wind is going
through a more evolved (and larger) AGB shell populated with predominantly
wider, randomly distributed and inhomogeneous voids.
This therefore generates (mostly) large outflows/lobes, although small
structures can also be present.
The waist often described in starfish nebulae can therefore be explained
as being the condensation of the nebular material at the base of the voids
in the filamentary shell. We note that in this scenario, there is no need
to invoke a particular launching mechanism or precession as it would be
the case in other bipolar or multipolar PNe: the wind-shell interaction,
owing the evolving filamentary nature of the latter, seems to work
appropriately.

\section*{Acknowledgments}

We would like to thank the reviewer for her/his thorough review and appreciate the comments and suggestions made.
LS acknowledges support from PAPIIT grant IA-101316 (Mexico).
GR-L acknowledges support from CONACYT, CGCI, PRODEP and SEP (Mexico).
MAG acknowledges support of the grant AYA 2014-57280-P, co-funded with
FEDER funds. SZ acknowledges support from the UNAM-ITE collaboration agreement 1500-479-3-V-04.
We thank the daytime and night support staff at the OAN-SPM for facilitating and helping obtain our observations   especially to Mr Gustavo  Melgoza-Kennedy,  our telescope  operator, for his assistance during observations, as well as the CATT for time allocation. This article is based upon observations carried out at the Observatorio Astron\'omico Nacional on the Sierra San Pedro M\'artir (OAN-SPM), Baja California, M\'exico. The data presented here were obtained in part with ALFOSC, which is provided by the Instituto de Astrofisica de Andalucia (IAA) under a joint agreement with the University of Copenhagen and NOTSA. We acknowledge the Instituto de Astrofisica de Canarias for the use of the filters. This research has made use of the NASA/IPAC Infrared Science Archive, which is operated by the Jet Propulsion Laboratory, California Institute of Technology, under contract with the National Aeronautics and Space Administration. We have also used archival observations made with the NASA/ESA Hubble Space Telescope, and obtained from the Hubble Legacy Archive, which is a collaboration between the Space Telescope Science Institute (STScI/NASA), the Space Telescope European Coordinating Facility (ST-ECF/ESA) and the Canadian Astronomy Data Centre (CADC/NRC/CSA). {\sc IRAF} is distributed by the National Optical Astronomy Observatory, which is operated by the Association of Universities for Research in Astronomy (AURA) under a cooperative agreement with the National Science Foundation.

\bibliographystyle{mn2e}

\bibliography{sabin_pc22}

\bsp

\appendix
\section{Observed and synthetic spectra of PC\,22}

\begin{figure*}
\begin{center}
\vspace{3cm}
\includegraphics[width=9cm]{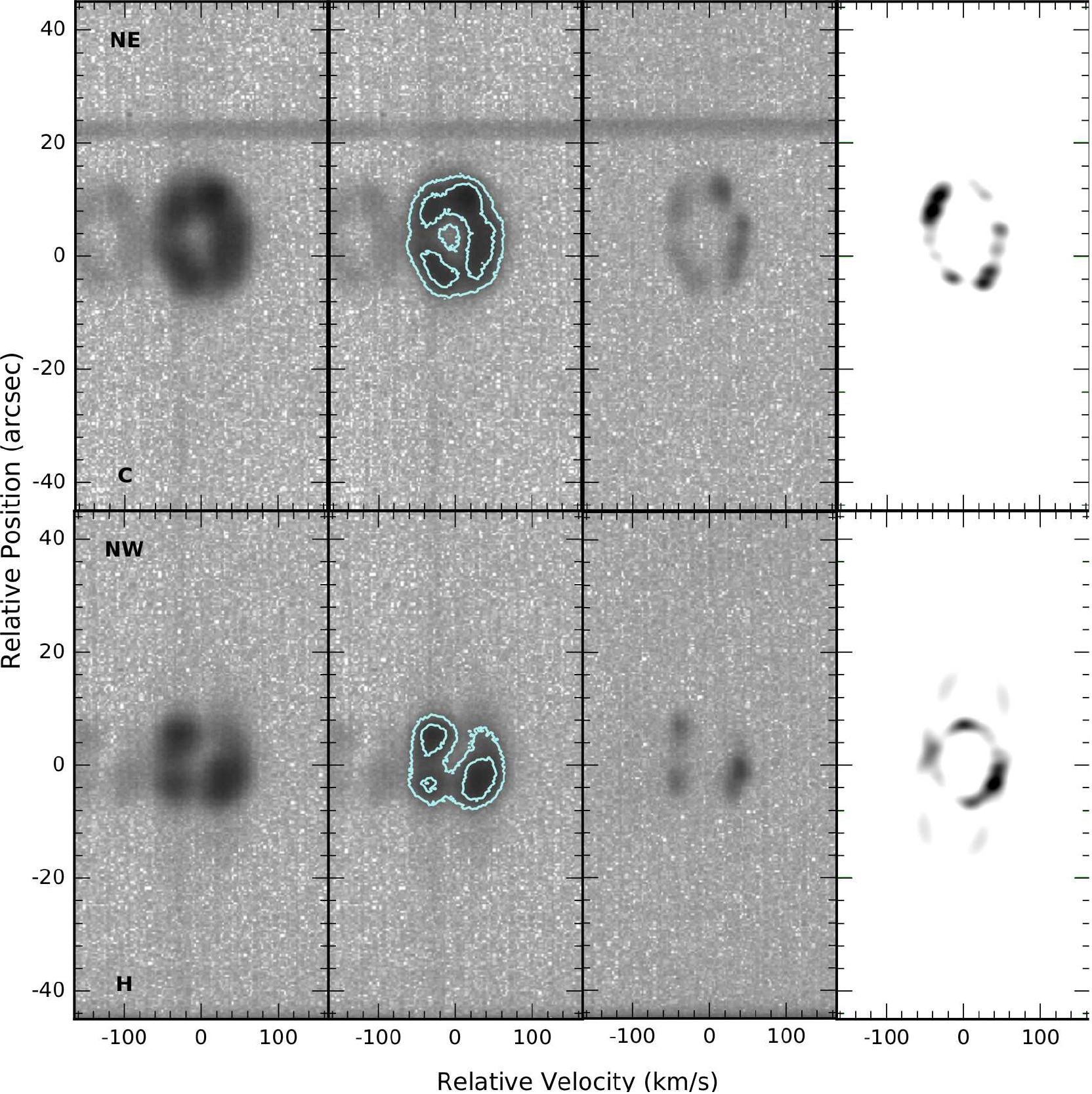}
\caption{From left to right: MES [O\,{\sc iii}], [O\,{\sc iii}]+contours, H$\alpha$  and [O\,{\sc iii}] line {\sc shape} modelling for the slits C (top) and H (bottom)}
\label{app1}
\end{center}
\end{figure*}

\begin{figure*}
\begin{center}
\vspace{2cm}
\includegraphics[width=14cm]{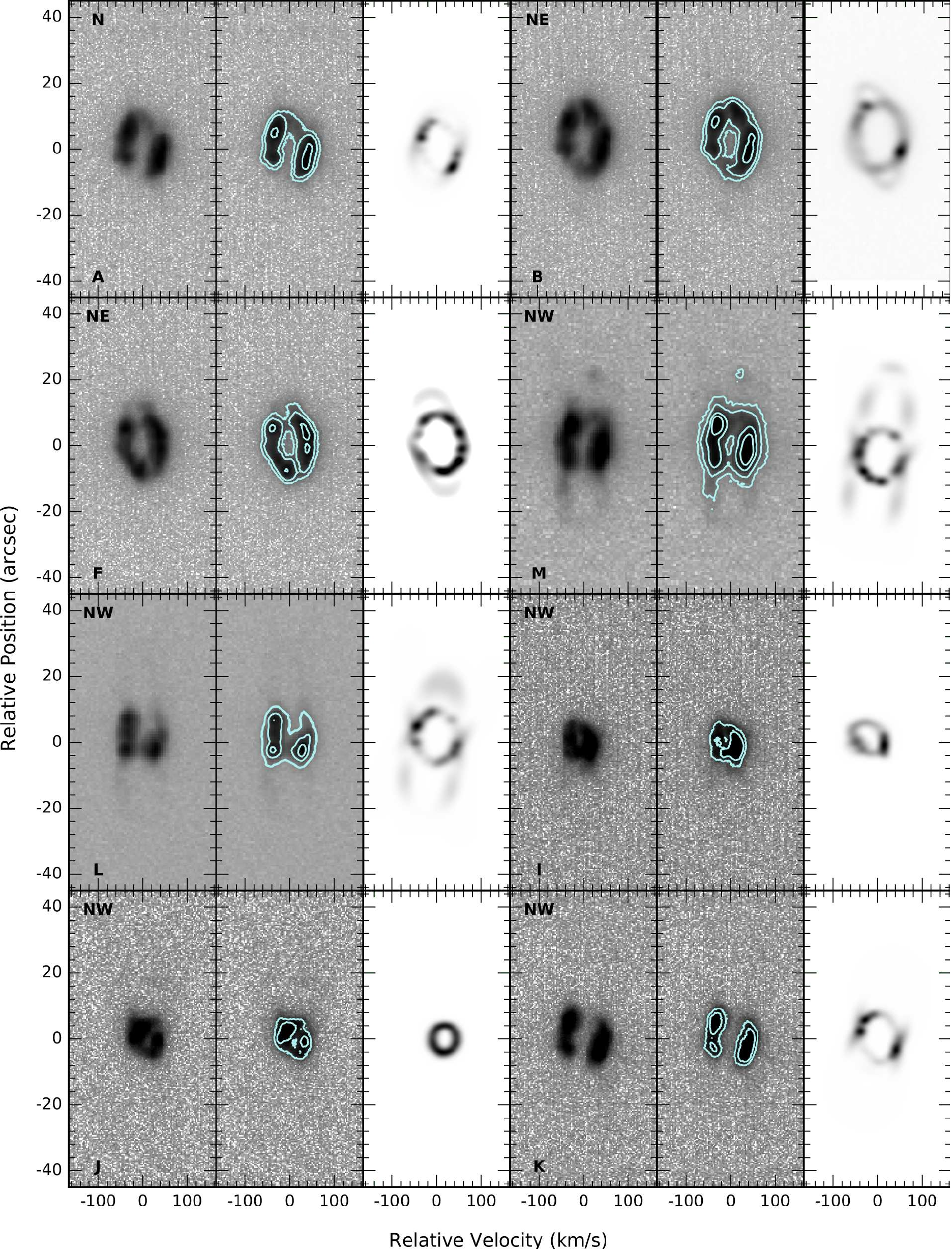}
\caption{MES [O\,{\sc iii}], [O\,{\sc iii}]+contours and [O\,{\sc iii}] line {\sc shape} modelling for the slits A,B,F,M,L,I,J and K.}
\label{app2}
\end{center}
\end{figure*}


\end{document}